\documentclass[aps,prb,reprint,floatfix,superscriptaddress]{revtex4-1}

\usepackage{amsmath,amsfonts,amssymb} %
\usepackage{mathtools} %
\usepackage{wasysym} %
\usepackage{bm} %
\usepackage[bbgreekl]{mathbbol} %
\usepackage{graphicx} %
\usepackage[dvipsnames,table]{xcolor} %
\usepackage{tabularx} %
\usepackage[acronym]{glossaries} %
\usepackage{braket} %
\usepackage{multirow} 
\usepackage{fancyhdr} %
\usepackage{microtype} %
\usepackage{natmove}

\usepackage[byname]{smartref} %
\usepackage[breaklinks, %
colorlinks, linkcolor=Blue, citecolor=Blue, urlcolor=Blue]{hyperref} %
\usepackage[version=3]{mhchem}
\AtBeginDocument{\usepackage{booktabs}} 
\newcommand{\eq}[1]{Eq.~(\ref{#1})} %
\newcommand{\fig}[1]{Fig.~\ref{#1}} %
\newcommand{\figs}[1]{Figs.~\ref{#1}} %
\def\be{\begin{equation}} %
\def\ee{\end{equation}} %

\newcommand{\CR}[1]{\hat a^{\dagger}_{#1}}
\newcommand{\AN}[1]{\hat a_{#1}}

\newcommand{\EE}[2]{\hat E_{#2}^{#1}}

\newcommand{\bea}{\begin{eqnarray}}
\newcommand{\eea}{\end{eqnarray}}
\newcommand{\LA}[1]{\mathfrak{#1}}

\newcommand{\HC}{\hat C}
\newcommand{\HA}[1]{\hat A_{#1}}

\begin{document}

\title{Cartan sub-algebra approach to efficient measurements of quantum observables}

\author{Tzu-Ching Yen}
\affiliation{Chemical Physics Theory Group, Department of Chemistry,
  University of Toronto, Toronto, Ontario, M5S 3H6, Canada}

\author{Artur F. Izmaylov}
\email{artur.izmaylov@utoronto.ca}
\affiliation{Department of Physical and Environmental Sciences,
  University of Toronto Scarborough, Toronto, Ontario, M1C 1A4,
  Canada}
\affiliation{Chemical Physics Theory Group, Department of Chemistry,
  University of Toronto, Toronto, Ontario, M5S 3H6, Canada}

\begin{abstract}
An arbitrary operator corresponding to a physical observable cannot be measured in a single measurement on currently 
available quantum hardware. 
To obtain the expectation value of the observable, one needs to partition its operator
to measurable fragments. 
%The number of such fragments can grow quite rapidly with the size of the system, thus, finding  
%grouping techniques minimizing the number of fragments is crucial for efficient extraction of information from the quantum system.   
However, the observable and its fragments generally do not share any eigenstates, 
and thus the number of measurements needed to obtain the expectation value of the observable can grow rapidly 
even when the wavefunction prepared is close to an eigenstate of the observable.
We provide a unified Lie algebraic framework for developing efficient measurement schemes for quantum observables,  
it is based on two elements: 1) embedding the observable operator in a Lie algebra and 
2) transforming Lie algebra elements into those of a Cartan sub-algebra (CSA) using unitary operators. % from the corresponding Lie group. 
The CSA plays the central role because all its elements are mutually commutative and thus can be measured simultaneously. 
We illustrate the framework on measuring expectation values of Hamiltonians appearing in the Variational Quantum Eigensolver 
approach to quantum chemistry. The CSA approach puts many recently proposed methods for the measurement optimization within a single 
framework, and allows one not only to reduce the number of measurable fragments but also the total number of measurements. 
\end{abstract}

\date{\today}
\maketitle

\section{Introduction} 

In digital quantum computing, one prepares a wavefunction of the simulated quantum system and any property of interest 
needs to be physically measured to obtain estimates that constitute the result of the computation. 
One popular example of this approach is the Variational 
Quantum Eigensolver (VQE)\cite{Peruzzo2014} that is used to solve an eigenvalue problem for the Hamiltonian of interest.
Note that in contrast to analogue quantum computing,\cite{cirac:2012,cirac:2018} the Hamiltonian 
of interest is not encoded in VQE but rather its expectation value is measured. The main problem 
of this measurement setup is that the entire Hamiltonian cannot be measured in a single measurement.
This makes efficient partitioning of operators to measurable components one of the most important problems 
of digital quantum computing. 

Any quantum observable is represented by an operator in some mathematical form. 
To obtain the expectation value of this operator in digital quantum computing, one needs to prepare the quantum 
system in a particular state corresponding to a wavefunction in the Hilbert space of $N$ qubits ($\ket{\Psi}$) and 
to represent the operator of interest $\hat O$ in that %$N$-qubit 
Hilbert space
\bea\label{eq:Oq}  
\hat O = \sum_n c_n \hat P_n, ~\hat P_n = \otimes_{k=1}^N \hat \sigma_k ,
\eea
where $c_n$ are numeric coefficients, and $\hat P_n$ are tensor products of 
single-qubit operators $\hat \sigma_k$, which are either Pauli spin operators $\hat x_k, \hat y_k, \hat z_k$
or the identity $\hat 1_k$. 

\begin{figure}[h!]
  \includegraphics[width=0.9\columnwidth ]{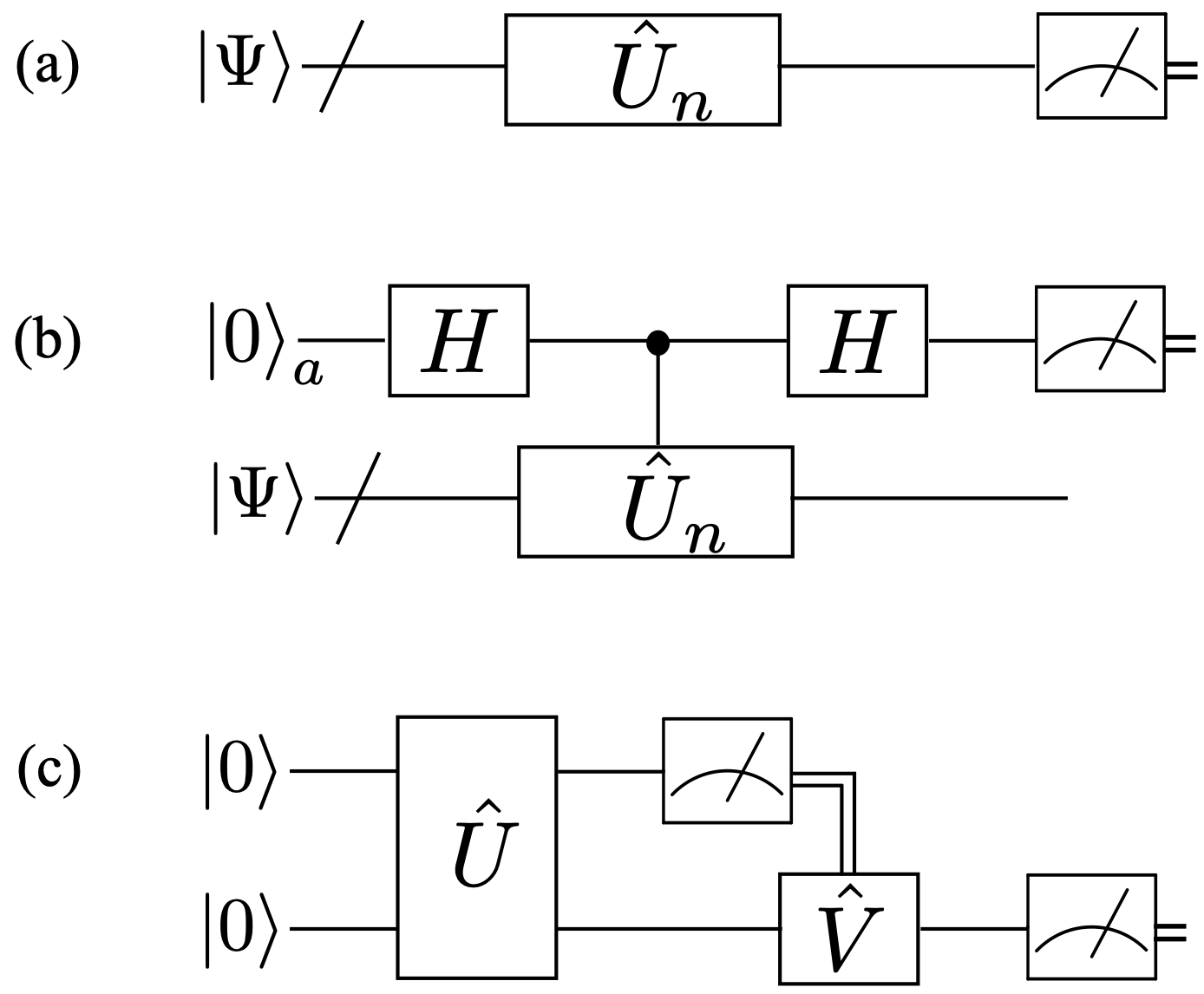}%, angle=90]{msch}
  \caption{Various measurement schemes involving unitary transformations: 
  (a) measuring $\bra{\Psi}\hat U_n^\dagger \hat z_k \hat U_n\ket{\Psi}$; (b) measuring 
  ${\rm Re} \bra{\Psi} \hat U_n \ket{\Psi}$ as $\pm(1-2p_{\pm 1})$, where 
  $p_{\pm 1}$ is the probability of obtaining $\pm 1$ in the auxiliary qubit measurement (here, $H$ is the Hadamard gate);
  (c) two-qubit example of a feed-forward scheme where a result of the first qubit measurement defines 
  application of the $\hat V$ transformation for the second qubit.}
  \label{fig:msch}
\end{figure}

Digital quantum computers can measure only single qubit polarization along the $z$-axis. 
This allows one to measure straightforwardly only $2^N-1$ operators for $N$ qubits: $\hat z_i$ and all 
possible products $\hat z_i\otimes\hat z_j ... \otimes\hat z_k$.  
An arbitrary operator $\hat O$ can be expressed as a linear combination of such products only after applying some 
multi-qubit unitary transformation $\hat U$:
\bea\label{eq:UzU}
\hat O = \hat U^\dagger \left[ \sum_i a_i \hat z_i + \sum_{ij} b_{ij} \hat z_i \hat z_j + ...\right] \hat U,
\eea
where $a_i,b_{ij},...$ are some constants. 
The problem of obtaining $\hat U$ is equivalent to solving the eigenvalue problem for $\hat O$, 
and thus is hard to solve in general.  
However, one can take advantage of additivity of the $\hat O$ expectation value with respect 
to $\hat O$ components.  %of  measuring 
A feasible alternative to representation in \eq{eq:UzU} is partitioning $\hat O$ into fragments 
$\hat O = \sum_n \hat O_n$, where each $\hat O_n$ can be written as \eq{eq:UzU} and 
has corresponding ``diagonalizing" $\hat U_n$.    
This raises a question: how to select such fragments $\hat O_n$? Two main requirements for these fragments 
are that there is not a large number of them required to represent the whole $\hat O$, and that $\hat U_n$ is not difficult 
to obtain using a classical computer. Having these fragments allows one to measure the expectation value of $\hat O$
on a wavefunction $\ket{\Psi}$ as in \fig{fig:msch}(a). The overall efficiency of the $\hat O$ measurement will depend
on how many measurements each fragment will require to obtain accurate estimates of their expectation values. This 
consideration makes fragment variances, $\langle\hat O_n^2\rangle-\langle\hat O_n\rangle^2$, to be key quantities for 
efficient measurement.  

There are multiple ways to choose $\hat O_n$'s considering the two conditions. For example, one can try to find exactly 
solvable models within $\hat O$ as $\hat O_n$ fragments, but some models are not straightforward to find within $\hat O$,
and their sizes may be small. Also, the sum of two exactly solvable models may not necessarily be another exactly solvable model, which prevents combining such models into larger fragments. 
One of the most productive approaches 
so far has been trying to increase the size of $\hat O_n$ by collecting all terms of $\hat O$ that mutually commute. 
The condition of mutual commutativity allows all commuting operators in each $\hat O_n$ to be simultaneously diagonalizable, or to have one unitary operator, $\hat U_n$, transforming all of them into all $\hat z_i$ form as in \eq{eq:UzU}.
This idea has been applied to the electronic Hamiltonian in both 
qubit\cite{Verteletskyi:2020do,Kandala:2017/nature/242,Yen2019b,MoscaA,ChicagoA,crawford2019efficient} and 
fermionic\cite{huggins2019efficient,BonetMonroig:2019wv} operator forms. It was found that working in the fermionic 
representation provided superior results compared to those obtained in the qubit case.\cite{huggins2019efficient,BonetMonroig:2019wv} 
Even though all these methods stem from the same idea of finding commuting terms, there was no general framework 
that would encompass all mentioned methods developed so far. This work will provide such a framework with solid 
foundations in theory of Lie algebras. Moreover, the introduced framework allows us to propose an extension
of previous approaches that results in lower numbers of measurements required to accurately obtain expectation value.

Alternative to a simple measurement scheme presented in \fig{fig:msch}(a), there were few more sophisticated 
schemes put forward for efficient measurement, see \figs{fig:msch}(b) and \ref{fig:msch}(c). 
The scheme on \fig{fig:msch}(b) exploits 
partitioning of the operator to a linear combination of unitaries that can be measured indirectly via measuring the 
orientation of an auxiliary qubit (the Hadamard test).\cite{Izmaylov2019unitary} This scheme requires efficient partitioning of the operator to 
unitary components and implementing them in the controlled form. Such controlled unitaries are generally more expensive in terms of 
two-qubit gates than their non-controlled counterparts. The scheme on \fig{fig:msch}(c) uses
results of previous measurements to adjust next measurements, this is a so-called feed-forward scheme. Recently, it 
became available not only in labs\cite{Prevedel:2007ca} but also in an actual quantum computer produced by Honeywell.\cite{Pino:2020uo} 
Efficiency of this scheme depends on partitioning of the operator into measurable fragments. Even though the definition 
of such fragments have been introduced,\cite{Izmaylov2019revising} there is still no systematic procedure to do 
the optimal partitioning of an arbitrary operator to such fragments.

Yet another class of measurement techniques that appeared recently are methods based on the shadow 
tomography.\cite{aaronson2018shadow,Huang_2020,Rubin_ST:2021}
It is based on stochastic sampling of unitary rotations $\hat U_n$ for optimal measurement of an operator set $\hat O_n$. 
Even though this technique was used for measurement of electronic Hamiltonians recently,\cite{hadfield2021adaptive,hadfield2020measurements,Robert_D:2021} 
its careful comparison with deterministic partitions developed here goes beyond the scope of the current 
work and will be reported elsewhere.\cite{Overlapping:2021}

The rest of the paper is organized as follows. 
Section II A shows how a quantum observable operator $\hat O$ can be expressed using elements of different Lie algebras. 
Section II B details the role of the CSA for measurement of $\hat O$ and CSA decomposition. 
Section III illustrates how application of the CSA decomposition to the fermionic and qubit algebras connects several 
existing approaches, 
%and gives rise to a new improved scheme, 
and Section IV proposes schemes improving CSA decompositions in the number of measurements. 
Section V assesses the new approach and several existing methods on a set of molecular systems 
(H$_2$, LiH, BeH$_2$, H$_2$O, NH$_3$, and N$_2$). 
Section VI summarizes the main results and provides concluding remarks.

\section{Theory} \label{sec:T}

\subsection{Operator embeddings}

Any operator in quantum mechanics can be written as a polynomial expression of some elementary operators $\{\HA{k}\}$
\bea \label{eq:O}
\hat O = \sum_{k} c_k \hat A_k + \sum_{kk'} d_{kk'} \hat A_k \hat A_{k'} +...,
\eea
where $\{\HA{k}\}$ form a Lie algebra $\mathcal{A}$ with respect to the commutation operation
\bea
[\hat A_i,\hat A_j] = \sum_k \xi_{ij}^{(k)} \hat A_k, ~\hat A_i,\hat A_j, \hat A_k \in \mathcal{A}
\eea
here $\xi_{ij}^{(k)} $ are so-called structural constants from the number field $\mathbb{K}$.\cite{Gilmore:2008} 
Note that products like $\hat A_k \hat A_{k'}$
 and higher powers of $\mathcal{A}$ elements do not generally belong to the Lie algebra $\mathcal{A}$, instead they are part of 
 a universal enveloping algebra (UEA), $\mathcal{E_A}$, which is built as a direct sum of tensor powers of the Lie algebra 
\bea
\mathcal{E_A} = \mathbb{K}\oplus\mathcal{A}\oplus(\mathcal{A}\otimes\mathcal{A})\oplus ...,
\eea
where the Lie bracket operation is equivalent to the commutator. 
Thus, any operator in quantum mechanics is an element of some UEA. 

Any operator for a physical observable is self-adjoint (i.e. hermitian), which allows us to only consider 
compact Lie algebras (i.e. all generators are hermitian) for expressing such operators. Any compact 
Lie algebra can be expressed as a direct sum of abelian and semisimple Lie algebras.\cite{Barut_p17} For considering 
the measurement problem within compact Lie algebras, it is convenient to introduce a notion of the Cartan 
sub-algebra $\mathcal{C}\subset \mathcal{A}$, which in this case, is a maximal abelian sub-algebra.
\footnote{All commuting elements of the semisimple Lie algebra must be ad-diagonalizable.} 
We will denote elements of $\mathcal{C}$ as $\hat C_k$'s.  
The UEA constructed from $\mathcal{C}$, $\mathcal{E_C}$, is abelian as well.
Thus, in principle, all elements of $\mathcal{E_C}$ can be measured simultaneously. In practice,
there is need for unitary transformations that transform $\mathcal{E_C}$ elements of a particular CSA
into those of all $\hat z$ qubit operators. For all algebras discussed in this work these transformations
involve only standard fermion-qubit mappings.  

One operator can be written as an element of different UEAs, where different UEAs are built from different Lie algebras. 
We will refer to expressions of the same operator using different Lie algebras as different embeddings. Various 
embeddings of the electronic Hamiltonian are illustrated in Appendix A.

\subsection{Cartan sub-algebra approach}
We will illustrate how elements of $\mathcal{E_C}$ can be used for measuring operator $\hat O$ 
that can be written using elements of UEA of a Lie algebra $\mathcal{A}$ 
up to quadratic terms, $\HA{k}\HA{k'}$ in \eq{eq:O}.  This case can be easily generalized 
to operators that contain higher powers of $\HA{k}$. Also, we generally assume that 
$\HA{k}\HA{k'}$ do not belong to the Lie algebra $\mathcal{A}$.
To use elements of $\mathcal{E_C}$ we need to find a minimum number of unitary transformations that allow us to present 
the operator of interest as  
\bea\label{eq:CO}
\hat O = \sum_{\alpha=1}^M \hat U_{\alpha}^\dagger \left [ \sum_l^{|\mathcal{C}|} \lambda_l^{(1,\alpha)} \HC_l + \sum_{ll'}^{|\mathcal{C}|} \lambda_{ll'}^{(2,\alpha)} 
\HC_l \HC_{l'} \right ]\hat U_{\alpha},
\eea
where $\HC_l\in\mathcal{C}$, $|\mathcal{C}|$ is the CSA size, $\lambda_l^{(1,\alpha)}$ and $\lambda_{ll'}^{(2,\alpha)}$ are some tensors. 

What are possible candidates for $\hat U_{\alpha}$? Clearly, they should depend on $\hat A_k$'s and do not create complicated expressions 
when act on CSA elements. In this work we will consider two constructions of $\hat U_{\alpha}$'s, but generally one can search 
for other ways to construct $\hat U_{\alpha}$'s as functions of $\hat A_k$'s. The main guiding principle in this search can be a requirement that
 $\hat U_{\alpha}$ transformation of any element of $\mathcal{E_C}$ produces a low-degree polynomial number of terms from $\mathcal{E}$. 
 
 \subsubsection{Lie group unitaries} 

The first approach for $\hat U_{\alpha}$'s is to take the elements of the corresponding Lie group 
\bea\label{eq:Ua}
\hat U_{\alpha} = \exp\left[ i\sum_k^{|\mathcal{A}|} \hat A_k \theta_k^{(\alpha)} \right],
\eea
where $\theta_k^{(\alpha)} \in \mathbb{R}$, $|\mathcal{A}|$ is the Lie algebra size, and $\hat A_k$'s are assumed to be hermitian. 
Due to the closure in $\mathcal{A}$ we have
\bea\label{eq:tran}
\hat U_{\alpha}^\dagger \left [ \sum_k^{|\mathcal{A}|} c_k \hat A_k \right ]\hat U_{\alpha} = \sum_k^{|\mathcal{A}|} c_k^{(\alpha)} \hat A_k, 
\eea
where $c_k^{(\alpha)}$ are some constants. Essentially, the Lie group elements $\hat U_{\alpha}$ provide a set of automorphisms 
for the corresponding Lie algebra. 
Moreover, according to the maximal tori theorem 
for compact groups (all algebras involved in the Hamiltonian embeddings correspond to compact groups),\cite{Hall:MTT} 
it is guaranteed that there exists a choice of $\theta_k^{(\alpha)}=\theta_k$ in $\hat U_{\alpha} = \hat U (\theta_k)$ satisfying
\bea\label{eq:MTT}
\hat U \left [ \sum_k^{|\mathcal{A}|} c_k \hat A_k \right ]\hat U^\dagger = 
\sum_l^{|\mathcal{C}|} \lambda_l^{(1)} \HC_l 
\eea
for any values of $c_k$. The maximal tori theorem provides a basis for finding a single $\hat U$ that transforms 
the linear part of $\hat O$ into a linear combination of the CSA terms. Amplitudes $\boldsymbol{\theta}=\{\theta_k\}$ 
for this $\hat U$ can be found numerically by solving the system of equations
\bea
 c_k &=& \sum_l^{|\mathcal{C}|} \lambda_{l}^{(1)} c_k^{(l)} (\boldsymbol{\theta}), 
\eea
where $c_k^{(l)} (\boldsymbol{\theta})$ are some functions whose explicit form depends on the Lie algebra.
Thus, in what follows we can focus 
on representation of the quadratic part of $\hat O$, $\hat O^{(2)}$.  

Since $\hat U_\alpha$'s in \eq{eq:Ua} do not change the power of $\hat A_k$'s after transformation \eqref{eq:tran},  
one can find $\theta_l^{(\alpha)}$, and $\lambda_{ll'}^{(2,\alpha)}$ by 
minimization of the difference between the quadratic parts of Eqs.~\eqref{eq:O} and \eqref{eq:CO}. 
To facilitate the process it is useful to introduce the appropriate basis in the UEA. Such bases are given by the 
Poincare--Birkhoff--Witt theorem:\cite{Barut:1980} 
\bea
&basis ~1&:~ \left\{1,\HA{k_1},\HA{k_1}\HA{k_2}\right\}, \\
&basis ~2&:~\left\{1,\HA{k_1},(\HA{k_1}\HA{k_2}+\HA{k_2}\HA{k_1})/2\right\}, 
\eea
where $k_2 \le k_1 =1,...,|\mathcal{A}|$. Note that these bases can be continued to the higher polynomial functions 
of $\HA{k}$'s but we do not need them beyond the quadratic terms.   

Both representations of $\hat O$ [Eqs.~\eqref{eq:O} and \eqref{eq:CO}] 
can be transformed to a linear combination of the basis elements. Symmetric 
{\it basis 2} is somewhat simpler to work with, and thus, we will use it here denoting 
$\{\HA{k_1}\HA{k_2}\}_S = (\HA{k_1}\HA{k_2}+\HA{k_2}\HA{k_1})/2$.
  Assuming hermiticity of $d_{kk'}$, the quadratic part of \eq{eq:O} transforms to
\bea\label{eq:qSO}
\hat O^{(2)} =  \sum_{k\ge k'}^{|\mathcal{A}|} d_{kk'}(2-\delta_{kk'}) 
\{\hat A_{k} \hat A_{k'}\}_S.
\eea
The $\hat O^{(2)}$ part of \eq{eq:CO} can be written as
\bea
\hat O^{(2)} &=& \sum_{\alpha=1}^M \sum_{ll'}^{|\mathcal{C}|} \lambda_{ll'}^{(2,\alpha)} 
[\hat U_{\alpha}^\dagger \HC_l \hat U_{\alpha}][\hat U_{\alpha}^\dagger \HC_{l'} \hat U_{\alpha}].
\eea 
Applying the unitary transformation to CSA elements
\bea
\hat U_{\alpha}^\dagger \HC_l \hat U_{\alpha} = \sum_k^{|\mathcal{A}|} c_k^{(l)} (\boldsymbol{\theta}^{\alpha}) \hat A_{k}
\eea
leads to  
\bea
\hat O^{(2)} &=& \sum_{k\ge k'}^{|\mathcal{A}|} \left[\sum_{\alpha=1}^M \sum_{ll'}^{|\mathcal{C}|} \lambda_{ll'}^{(2,\alpha)} c_k^{(l)} (\boldsymbol{\theta}^{\alpha})c_{k'}^{(l')} (\boldsymbol{\theta}^{\alpha}) \right] \notag\\ \label{eq:qSO2}
&& (2-\delta_{kk'}) \{\hat A_{k} \hat A_{k'}\}_S .
\eea
Term-wise comparison of Eqs.~\eqref{eq:qSO} and \eqref{eq:qSO2} gives equations 
on $\boldsymbol{\lambda}^{(2,\alpha)}$ and $\boldsymbol{\theta}^{(\alpha)}$
 \bea \label{eq:djj}
 d_{kk'} &=& \sum_{\alpha=1}^M \sum_{ll'}^{|\mathcal{C}|} \lambda_{ll'}^{(2,\alpha)} c_k^{(l)} (\boldsymbol{\theta}^{(\alpha)})c_{k'}^{(l')} (\boldsymbol{\theta}^{(\alpha)}),
 \eea
where the only functions to derive are $c_k^{(l)} (\boldsymbol{\theta}^{\alpha})$. 

This consideration can be extended to higher powers of Lie algebra elements beyond quadratic. 
However, for such extensions %the operator embeddings that involve higher powers of $\hat A_k$'s,
the number of equations will grow exponentially with the algebraic degree and become computationally overwhelming. 
%This happens because Lie group elements are automorphisms of the Lie algebra that 
%conserve the powers of $\hat A_k$'s in the $\hat O$ expression. 

 \subsubsection{Number of terms conserving unitaries}

An alternative that is more efficient for cases with 
higher powers of $\hat A_k$'s is to use $\hat U_\alpha$'s that conserve the number of terms in the transformation:
\bea\label{eq:U2}
\hat U_\alpha^\dagger \left(\prod_i \hat C_{k_i} \right)\hat U_\alpha =  \prod_j \hat A_{k_j}. 
\eea
The functional form of $\hat U_\alpha=\hat U_\alpha(\{\hat A_k\})$ 
depends on associative algebraic properties of the used UEA, and cannot be given explicitly 
for a general UEA. For example, in the qubit embedding, every element of the Lie algebra is involutory, 
$\hat A_k \hat A_k = \hat 1$, which leads to $\hat U_\alpha$'s to be elements of the Clifford group.    

Since all elements of $\mathcal{E_C}$ commute among themselves, 
the resulting products of $\hat A_{k_j}$ in \eq{eq:U2} are also commutative. Thus, 
practically, to find a linear combination of operator terms that can be treated by a single $\hat U_\alpha$,
one needs to group all mutually commuting terms. Generally, the commutation relation between operator terms
can be represented by a graph whose edges connect commuting terms represented by vertices. 
Partitioning of the $\hat O$ expression to mutually commuting groups can be done in various ways, 
but finding the optimal partitioning to a minimum number of such groups 
is a standard NP-hard problem in graph theory, the minimum clique cover problem. 
There are many heuristic polynomial algorithms to solve this problem.\cite{Verteletskyi:2020do}

\section{Applications}\label{sec:app}

Here we will illustrate how novel algorithms and several previously developed methods 
can be derived from the CSA framework applied to the electronic Hamiltonian $\hat H_e$ as an example of 
a many-body operator $\hat O$. Fermionic and qubit embeddings will be used to illustrate the CSA decomposition of $\hat H_e$.
As shown in Appendix A, the fermionic embeddings contain low powers of Lie algebra elements and thus one can expect more benefits 
from using the Lie group unitaries [\eq{eq:Ua}], while the qubit embedding has the involutory associative multiplication property for the algebra elements
and thus benefits more from the number of term conserving unitaries [\eq{eq:U2}].   

\subsection{Fermionic algebras}

To apply the Lie group unitaries [\eq{eq:Ua}] one needs operator embeddings involving compact Lie algebras. 
Even though we use generators of the non-compact $\LA{gl}(N)$ algebra, $\EE{p}{q}$'s, this does not create 
any problem because in what follows $\EE{p}{q}$'s always appear as linear combinations $i(\EE{p}{q} - \EE{q}{p})$ and 
$(\EE{p}{q} + \EE{q}{p})$, which are generators of the compact Lie algebra $\LA{u}(N)$. 
The origin of this compactness is the hermiticity of the system Hamiltonian 
that allows to rewrite the Hamiltonian in $\LA{u}(N)$ generators. 
The CSA of $\LA{gl}(N)$ and $\LA{u}(N)$ are the same: $N$ elements $\EE{p}{p}$. 

$\EE{p}{q}$'s are isomorphic to real $N$ by $N$ matrices $(E_q^p)_{mn}=\delta_{mp}\delta_{nq}$, 
and there are $N^2$ such elements. This faithful representation of $\EE{p}{q}$ makes search for 
$\hat U_{\alpha}$ in \eq{eq:MTT} equivalent to a simple diagonalization of the hermitian matrix, 
$h_{pq}$. 

Unfortunately, the two-electron part of the fermionic Hamiltonian cannot be treated as easily as the one-electron part.
Products $\EE{p}{q}\EE{r}{s}$ do not form a Lie algebra, and thus their representation 
as tensor products of pairs of $N$ by $N$ matrices do not lead to unitary transformations that we are looking for. 
However, diagonalization of 4-index tensors $g_{pq,rs}$ can provide some approximation to 
the CSA decomposition and is discussed as a Hamiltonian factorization.      

\subsubsection{Hamiltonian factorization}

Using the maximal tori theorem one can employ the following heuristic approach to obtaining the expansion for the 
quadratic part $\hat O^{(2)}$ in products of linear combinations of $\hat A_{k}$  
\bea\label{eq:Q}
\hat O^{(2)} &=& \sum_{\alpha=1}^{M'} \left( \sum_k^{|\mathcal{A}|} c_k^{(\alpha)} \hat A_k \right)^2 \\ \label{eq:CD}
&=& \sum_{k,k'}^{|\mathcal{A}|} \hat A_{k}\hat A_{k'} \sum_{\alpha=1}^{M'} [c_k^{(\alpha)}c_{k'}^{(\alpha)}]  \\
&=& \sum_{k,k'}^{|\mathcal{A}|} \tilde{d}_{kk'} \hat A_k\hat A_{k'},   
\eea
where $\tilde{d}_{kk'}$ is an analogue of $d_{kk'}$ from \eq{eq:O}. Thus, one can 
decompose $d_{kk'}$ as $\sum_{\alpha=1}^{M'} [c_k^{(\alpha)}c_{k'}^{(\alpha)}]$ and then find $\hat U_{\alpha}$ by applying the the maximal tori theorem for each 
linear combination $\sum_k c_k^{(\alpha)} \hat A_k$.
Yet, considering that the rank for $\lambda_{ll'}^{(2,\alpha)}$ in \eq{eq:CO} can be $|\mathcal{C}|$ (full rank) 
and that of $c_k^{(\alpha)}c_{k'}^{(\alpha)}$ for each $\alpha$ in \eq{eq:CD} is only 1, 
one can see that the representation of \eq{eq:CO} can be more compact than that of \eq{eq:CD}, 
$M<M'$. 

A heuristic treatment of the quadratic part of $\hat O$ by decomposing 
$d_{kk'}$ to $c_k^{(\alpha)}c_{k'}^{(\alpha)}$ in \eq{eq:CD} can be done using singular value decomposition (SVD). 
\citet{huggins2019efficient} do exactly that for the two-electron part of the fermionic Hamiltonian where 
$\{d_{kk'}\} = \{g_{pq,rs}\}$ and $\{\hat A_k \}= \{\EE{p}{q}\}$ (details are given in Appendix B).

\subsubsection{Full rank optimization}

Extension of the factorization approach to \eq{eq:CO} requires substituting SVD to a more general decomposition.
To arrive at \eq{eq:CO}, one can start with the symmetrized form of $\hat H_e^{(2)}$
\bea 
\hat H_e^{(2)} &=& \sum_{pqrs}^{N} {g}_{pq,rs} \{\EE{p}{q} \EE{r}{s} \}_S
\eea
where $g_{pq,rs} = g_{rs,pq}$ due to permutation symmetry in the two-electron integrals. 
The equivalent of \eq{eq:CO} for this Hamiltonian is 
\bea\notag
\hat H_e^{(2)} &=& \sum_{\alpha=1}^M 
\hat U_{\alpha}^\dagger \left [ \sum_{tu}^{N} \lambda_{tu}^{(2,\alpha)} 
\EE{t}{t} \EE{u}{u} \right ]\hat U_{\alpha} \\ \label{eq:fmf}
&&= \sum_{\alpha=1}^M \sum_{tu}^{N} \lambda_{tu}^{(2,\alpha)} 
[\hat U_{\alpha}^\dagger \EE{t}{t} \hat U_{\alpha}][\hat U_{\alpha}^\dagger \EE{u}{u} \hat U_{\alpha}],
\eea 
where
\bea
\hspace{-0.5cm} \hat U_{\alpha} = \exp\left[
    {\sum_{t>u}^N-i\theta_{tu}^{(\alpha)}(\EE{t}{u}+\EE{u}{t})}
    + {\phi_{tu}^{(\alpha)}(\EE{t}{u}-\EE{u}{t})}
  \right].
\eea	
There is a homomorphism of these unitary operators onto unitary matrices that are obtained by substituting 
the excitation operators $\EE{t}{u}$ by matrices $E^{t}_{u}$, which is the faithful representation of algebraic generators. 
This homomorphism allows us to perform the operator 
transformations in \eq{eq:fmf} 
\bea
\hat U_{\alpha}^\dagger \EE{t}{t} \hat U_{\alpha} = \sum_{pq} \EE{p}{q} c_{pq}^{(t)}(\boldsymbol{\theta}^{(\alpha)},\boldsymbol{\phi}^{(\alpha)})
\eea 
by substituting the operators with corresponding matrices. This substitution gives coefficients 
$c_{pq}^{(t)}(\boldsymbol{\theta}^{(\alpha)},\boldsymbol{\phi}^{(\alpha)})$ via linear algebra. 
Assembling the operator expression of \eq{eq:fmf} after the transformation of the CSA elements provides
\bea\notag
\hat H_e^{(2)} &=& \sum_{\alpha=1}^M \sum_{tu}^{N} \lambda_{tu}^{(2,\alpha)} 
c_{pq}^{(t)}(\boldsymbol{\theta}^{(\alpha)},\boldsymbol{\phi}^{(\alpha)}) c_{rs}^{(u)}(\boldsymbol{\theta}^{(\alpha)},\boldsymbol{\phi}^{(\alpha)}) \\
&&\times\{\EE{p}{q} \EE{r}{s} \}_S.
\eea
Using independence of the basis set elements  $\{\EE{p}{q} \EE{r}{s} \}_S$, the following system of algebraic equations can be 
written
\bea\notag
{g}_{pq,rs} &=& \sum_{\alpha=1}^M \sum_{tu}^{N} \lambda_{tu}^{(2,\alpha)} 
c_{pq}^{(t)}(\boldsymbol{\theta}^{(\alpha)},\boldsymbol{\phi}^{(\alpha)}) 
c_{rs}^{(u)}(\boldsymbol{\theta}^{(\alpha)},\boldsymbol{\phi}^{(\alpha)}) \\
&=&  \sum_{\alpha=1}^M f_{pq,rs}^{(\alpha)}(\boldsymbol{\lambda}^{(2,\alpha)},\boldsymbol{\theta}^{(\alpha)},\boldsymbol{\phi}^{(\alpha)}).
\label{eq:CSA_vector_diff}
\eea
Its solution is found via gradient minimization of the norm for the vector of differences between the right and left hand sides using 
 the Broyden-Fletcher-Goldfarb-Shanno (BFGS) algorithm.\cite{Book/Press:1992}
% To determine $M$, we start with $M=N$ and increase its value if the norm of the difference vector is higher than a threshold. 
%To expedite this procedure one can start optimization for higher $M$ values using as initial guesses results from lower $M$ optimizations.

\subsection{Qubit algebras}

There are two compact, semisimple qubit algebras that can be used for the CSA decomposition 
of the qubit counterpart of the electronic Hamiltonian, $\hat H_q$ (Appendix A): 
1) UEA of $\mathcal{S}=\oplus_{k=1}^N \LA{su}(2)_k$, $\mathcal{E_S}$, or 2) $\LA{so}(2^N)$. 
One advantage of $\mathcal{E_S}$ is a single-particle picture (elements of $\mathcal{S}$) 
that can define a class of computationally feasible unitaries $\hat U_\alpha$. 

\subsubsection{Single-qubit unitaries}

The Lie group constructed by exponentiating the $\mathcal{S}$ algebra 
consists of the following elements 
\bea
\hat U_{\rm QMF} = \prod_{k=1}^N e^{i\tau_k(\bar{n}_k,\bar{\sigma}_k)},
\eea
where $\tau_k$ is an amplitude, $\bar{n}_k$ is a unit vector on the Bloch sphere, and $\bar{\sigma}_k = (\hat x_k,\hat y_k,\hat z_k)$.
The fragments that can be measured after $\hat U_{\rm QMF}$ transformations 
are exactly solvable within the qubit mean-field (QMF) approach\cite{Ryabinkin:2018di} 
\bea\label{eq:QMF}
\hat H_{\rm QMF} &=&  \hat U_{\rm QMF}^\dagger\left[ \sum_k a_k \hat C_k^{(Z)} \right] \hat U_{\rm QMF},
\eea
where $\hat C_k^{(Z)}$ are elements from the directly measurable CSA: $\{1, \hat z_i, \hat z_i \hat z_j, ... \}$. 
One-qubit rotations in the qubit space do not translate to one-electron fermionic transformations as was shown.\cite{Ryabinkin:2018di} 
Therefore, the fragments in \eq{eq:QMF} are different from those in \eq{eq:fmf}. 
The latter are also exactly solvable but with the fermionic one-particle transformations. 
 
Identifying fragments of \eq{eq:QMF} is not straightforward and requires a tensor decomposition 
of the qubit Hamiltonian, $\hat H_q$. Note that since $\hat U_{\rm QMF}$ transformations are not necessarily in 
the Clifford group,\footnote{Clifford group consists of unitaries that transform a single Pauli product to another Pauli product.}
each of them can produce exponentially many Pauli products by transforming a single product. The
number of terms in the directly measurable CSA in \eq{eq:QMF} can be also exponentially large. 
Thus, even though the total number of $\hat H_q$ terms is $O(N^4)$, a
general tensor decomposition \cite{Hitchcock_CPdecomp, Kolda_CPdecomp} of $\hat H_q$ to fragments can have exponential 
computational cost and therefore is unfeasible on a classical computer.

To avoid exponential growth of terms, one can restrict $\hat U_{\rm QMF}$ to a Clifford subset of transformations, thus switching 
from the Lie group unitaries (\eq{eq:Ua}) to the unitaries conserving the number of terms (\eq{eq:U2}). 
Then, the only fragments of $\hat H_q$ that will be transformed into $\hat C_k^{(Z)}$ are the qubit-wise commuting (QWC) 
sets of Pauli products.\cite{Verteletskyi:2020do} A practical way to find all QWC sets of Pauli products 
was found through mapping this problems to the minimum clique cover problem for a graph representing qubit-wise 
commutativity relation in $\hat H_q$.

\subsubsection{Multi-qubit unitaries}

Alternative consideration of $\hat H_q$ within the $\LA{so}(2^N)$ algebra has an advantage that 
the maximal tori theorem guarantees existence of a single $\hat U\in Spin(2^N)$\footnote{$Spin(K)$ is a Lie group that originates 
from exponentiation of $\LA{so}(K)$ generators.} for the transformation 
\bea
\hat U \hat H_q \hat U^\dagger = \sum_k a_k \hat C_k^{(Z)}.
\eea
Unfortunately, this unitary requires an exponential number of algebra elements, $\hat U = \exp[i\sum_k c_k \hat P_k]$.  
Therefore, the exponential size of $\LA{so}(2^N)$ prevents us from taking advantage of the maximal tori theorem. 
For practical purpose, we restrict $\hat U$ to unitaries from the Clifford group that conserve the number of terms \eq{eq:U2}. 
In this case, a single $\hat U$ will not be sufficient and $\hat H_q$ decomposition involves several terms
\bea\notag
\hat H_q &=& \sum_\alpha \hat H_q^{(\alpha)} \\\label{eq:C1}
&=&\sum_\alpha\hat U_\alpha^\dagger \left[\sum_k a_k^{(\alpha)} \hat C_k^{(Z)} \right] \hat U_\alpha.
\eea
To find $\hat U_\alpha$'s one can use the main property of the Clifford group that any Pauli product is transformed 
by an element of the Clifford group to another Pauli product. This means that any fragment $\hat H_q^{(\alpha)}$ consists 
of commuting Pauli products because their Clifford unitary images $\hat C_k^{(Z)}$ are also commuting. 
This consideration connects \eq{eq:C1} with a partitioning method based on finding 
fully commuting (FC) sets described in Ref.~\citenum{Yen2019b}. FC sets are also found by heuristic solutions
of the minimum clique cover problem for a graph built for $\hat H_q$ using the commutativity relation. 

\section{Measurement Optimization}
% \section{Variance Optimization}

%Discuss our techniques to 1) distribution of number measurements based on HF variances, 
%2) reducing the variance sum by greedy and other heuristics. 

%The main idea to keep it general as much as possible across different algebras and to mention some peculiarities of qubit vs fermionic cases at the end. 

In this section we review the standard approach to computing the number of measurements $K$ needed to estimate the expectation value of the electronic Hamiltonian within error $\epsilon$. 
For a decomposition of the Hamiltonian 
\bea
  \hat H &=& \sum_{\alpha=1}^M \hat H^{(\alpha)} , 
  \label{eq:meas_decomp}
\eea 
where the expectation value of each fragment $\hat H^{(\alpha)}$ is sampled $K^{(\alpha)}$ times, the total number of measurements is $K = \sum_\alpha K^{(\alpha)}$. 
As shown previously,\cite{Rubin_2018,crawford2019efficient} 
the optimal choice of $K^{(\alpha)}$ is 
\bea 
  K^{(\alpha)} = \frac{1}{\epsilon^2} \sqrt{\text{Var}_{\Psi}(\hat H^{(\alpha)})} \sum_{\beta=1}^M \sqrt{\text{Var}_{\Psi}(\hat H^{(\beta)})}, 
\eea 
where $\epsilon$ is the desired accuracy, $\ket{\Psi}$ is the VQE 
trial wavefunction, and $\text{Var}_{\Psi}(\hat H^{(\alpha)}) = \bra{\Psi}{\hat H^{(\alpha)}}{}^2 \ket{\Psi}- \bra{\Psi}\hat H^{(\alpha)}\ket{\Psi}^2$ is the operator variance. 
Then the total number of measurements is 
\bea
  K  = \frac{1}{\epsilon^2}\left( \sum_{\alpha=1}^M \sqrt{\text{Var}_{\Psi}(\hat H^{(\alpha)})} \right)^2, 
  \label{eq:sum_sqrt}
\eea 
where the square of the expression in the brackets can be seen as the variance of the estimators for $\bra{\Psi} \hat H \ket{\Psi}$ originating from the decomposition in \eq{eq:meas_decomp}. 

There are two main heuristics one could employ to reduce $K$. 
First, it is generally favorable to have an uneven distribution of $\text{Var}(\hat H_\alpha)$ due to the square root functions.\cite{crawford2019efficient} 
One can achieve this by iteratively applying FRO with few $M$ until convergence, where at each iteration one finds the fragment that minimizes the norm of the remaining coefficients using the ``greedy" approach. This approach generally results in concentrated coefficients and thus larger variances in the first few fragments. 
Second, one can approximate $K$ with $K_\varphi$ computed using wavefunction $\ket{\varphi}$ calculated from classical methods (e.g., Hartree-Fock (HF)) and then search for partitioning that minimizes $K_\varphi$. 
% This can be easily done in the full-rank optimization (FRO) by minimizing instead the sum of $w K_\varphi$ and vector norm of differences between the left and right hand side of \eq{eq:CSA_vector_diff}, where $w$ is some appropriate weight. 
Based on these two approaches, we propose the following modifications of the 
full rank optimization (FRO) that reduce the total number of measurements: 

\paragraph{Greedy FRO (GFRO):} 
This algorithm iteratively applies the FRO with $M=1$. 
At the $\alpha$-th iteration, GFRO finds the $\boldsymbol{\lambda}^{(2,\alpha)},\boldsymbol{\theta}^{(\alpha)},\boldsymbol{\phi}^{(\alpha)}$ of \eq{eq:CSA_vector_diff} that minimizes \bea
  \left|g_{pq, rs} - \sum_{\beta=1}^{\alpha-1} f^{(\beta)}_{pq,rs} - f^{(\alpha)}_{pq, rs}\left(\boldsymbol{\lambda}^{(2,\alpha)},\boldsymbol{\theta}^{(\alpha)},\boldsymbol{\phi}^{(\alpha)}\right)\right|^2
\eea where parameters of $f^{(\beta)}_{pq,rs}$ were fixed from previous iterations. 

% {\bf Variance FRO (VFRO):} 

\paragraph{Variance-estimate Greedy FRO (VGFRO):} 
This algorithm also uses FRO with $M=1$ consecutively and fixes coefficients from previous iterations. 
Unlike GFRO, VGFRO minimizes \bea 
  \left| d_{pq,rs}^{(\alpha)} \right|^2 + w\text{Var}_{\varphi}(\hat H^{(\alpha)})
  \label{eq:IVFROcost}
\eea at the $\alpha$-th iteration, where \bea 
  d^{(\alpha)}_{pq,rs} &=& g_{pq, rs} - \sum_{\beta=1}^{\alpha} f^{(\beta)}_{pq,rs} \\
  \hat H^{(a)} &=& \sum_{pqrs}^N f^{(\alpha)}_{pq,rs} \{\EE{p}{q} \EE{r}{s} \}_S
\eea and $w$ is a fixed parameter. Note that one needs to choose a $w$ small enough to encourage the optimization to reduce $\left| d_{pq,rs}^{(\alpha)} \right|^2$ despite the second term of \eq{eq:IVFROcost}, which is minimized when $f^{(\alpha)}_{pq,rs} = 0$. 

In practice, the variances of the first few fragments obtained from GFROs are significantly larger than those of the fragments from subsequent iterations. Thus, for computational efficiency of VGFRO, 
we drop the second term of \eq{eq:IVFROcost} after first $\mu$ fragments. 
This is equivalent to performing GFRO on the remaining $d^{(\mu)}_{pq, rs}$ tensor. 

\paragraph{Qubit-based algorithms:} In qubit algebra, the greedy approach naturally leads to the Sorted Insertion (SI) algorithm, \cite{crawford2019efficient} which also groups the commuting Pauli products with large coefficients together in a few measurable fragments. 
However, further utilizing approximated variance in SI is not straightforward, since partitioning of the qubit Hamiltonian is not guided by gradients that are used to minimize the norm and variance in a continuous fashion in VGFRO. 
%Empirically, we applied the idea similar to VGFRO to qubit algebra, but found the resulting number of measurements required can increase or decrease with respect to that of SI depending on the value of $w$. 
%Unfortunately, the optimal choice of $w$ is generally not known a priori.  

\begin{table*}[!htbp]%
  \caption{Number of measurable groups provided by different methods for Hamiltonians of several molecular systems  
  [the number of spin-orbitals ($N$) and the total number of Pauli products in the qubit form (Total)]: 
  qubit-algebra methods based on qubit-wise and full commutativity (QWC and FC), fermionic-algebra methods based on 
  the SVD factorization (SVD)\cite{huggins2019efficient}, full rank optimization (FRO), greedy FRO (GFRO) and variance-estimate greedy FRO (VGFRO). The norm-1 accuracy for the fermionic-algebra methods is  
  $10^{-5}$.} 
  {
  %\begin{tabular*}{\columnwidth}{@{\extracolsep{\fill}} l l l l l l l l}
  \begin{tabularx}{\textwidth}{l @{\extracolsep{\fill}} ccccccccc}
    \toprule
   Systems & $N$ & Total & QWC-LF & QWC-SI & FC-SI & SVD & FRO & GFRO & VGFRO\\
    \midrule
    H$_2$  & 4  & 15   & 3   & 3 & 2  & 4  & 3 & 3  & 4\\
    LiH    & 12 & 631  & 142 & 155 &42 & 22 & 9 & 78 & 97\\
    BeH$_2$& 14 & 666  & 172 & 183 & 36 & 29 & 13 & 118& 129\\
    H$_2$O & 14 & 1086 & 313 & 334 & 50 & 29 & 11 & 119& 148\\
    NH$_3$ & 16 & 3609 & 1272& 1359 & 122& 37 & 13 & 187& 208\\
    N$_2$  & 20 & 2951 & 1177& 1254 & 74 & 52 & 20 & 351& 349\\
    \bottomrule
  \end{tabularx} 
  }
  \label{tab:ngroup_result}
\end{table*}

\begin{table*}[!htbp]%
  \setlength\tabcolsep{0pt}
  \caption{
    The number of spin-orbitals ($N$) and the resulting number of measurements ($\epsilon^2 K$) required to estimate energy expectation up to a standard deviation $\epsilon$.
    %from applying different methods to the electronic Hamiltonians of several molecular systems. 
    Fragment variances are calculated using the exact ground states of the systems' Hamiltonians. 
    %Qubit-algebra methods are based on qubit-wise and full commutativity (QWC and FC), 
    %whereas fermionic-algebra methods are based on 
    %the SVD factorization (SVD)\cite{huggins2019efficient} and iterative full-rank optimizations (IFROs). 
    %The norm-1 accuracy for the fermionic-algebra methods is $10^{-5}$.
  } 
  {\begin{tabular*}{\textwidth}{@{\extracolsep{\fill}}l cccccccc }
      \toprule
      %  \hline\hline
      Systems & N & QWC-LF & QWC-SI & FC-SI & SVD & FRO & GFRO & VGFRO \\
      % \hline 
      \midrule
      H$_2$   & 4 & 0.136 & 0.136  & 0.136 & 0.136 & 0.123 & 0.136 & 0.136 \\
      LiH     & 12& 5.84 & 2.09 & 0.882 & 3.16 & 28.5 & 2.71 & 2.26 \\
      BeH$_2$ & 14& 14.3 & 6.34 & 1.11  & 1.86 & 24.6 & 1.47 & 0.851\\ 
      H$_2$O  & 14& 116  & 48.6& 7.59    & 58.5 & 389 & 49.4 & 46.2 \\ 
      NH$_3$  & 16& 352  & 97.0& 18.8    & 58.1 & 471 & 47.0 & 42.2 \\
      N$_2$   & 20& 445  & 193& 8.83    & 10.5 & 441 & 8.62  & 5.27  \\
      \bottomrule
      %\hline\hline
      \end{tabular*} 
  }
  \label{tab:variance_result}
\end{table*}

\section{Results}

To assess the CSA decomposition techniques in fermionic and qubit algebras, 
we applied them to a set of Hamiltonians previously used to demonstrate performance of similar measurement techniques \cite{Verteletskyi:2020do, Izmaylov2019unitary,Yen2019b} (Tables~\ref{tab:ngroup_result} and \ref{tab:variance_result}).
Details of these Hamiltonians are provided in Appendix C. 
%Since one-electron terms can be measured in a single measurement (this is a consequence of the maximal tori theorem \eq{eq:MTT} for the $\LA{u}(N)$ algebra), 
%we excluded them from the electronic Hamiltonians. 
QWC fragments were obtained by the largest first (LF) heuristic \cite{Verteletskyi:2020do} and the sorted insertion (SI) algorithm.\cite{crawford2019efficient} Comparing the number of fragments in QWC-LF and QWC-SI
shows slight advantage of QWC-LF (Table~\ref{tab:ngroup_result}). 
However, the number of measurements is almost factor of 2 lower 
for large Hamiltonians when fragments are defined by QWC-SI (Table~\ref{tab:variance_result}). 
This shows a general trend that the number of fragments does not always correlate with the 
number of measurements. Also, this illustrates advantage of the SI algorithm for the qubit term grouping, therefore for the FC method we used the SI algorithm as well.  
%the latter of which finds the FC fragments using a heuristic similar to that of GFRO. 

All fermionic-algebra methods approximate the two-electron integral tensor with a finite accuracy, 
which we judiciously chose to be $10^{-5}$ in 1-norm of the difference between all $g_{pq, rs}$ 
before the symmetrized algebra generator products and their restored values. 
For the SVD factorization approach, the singular values arranged in the descending order
and their eigenvectors were used to reconstruct the $g_{pq, rs}$ matrix until the 1-norm threshold was satisfied.

%For IFROs, it was found that allowing complex rotation generators in $\hat U_\alpha$ and complex values of $\boldsymbol{\lambda}^{(2,\alpha)}$ 
% did not provide any advantage in reducing the number of terms ($M$). 
%was not necessary for decomposition of the two electron part of the electronic Hamiltonian. 
For the FRO based algorithms, it was found that real rotation generators in $\hat U_\alpha$ and real $\boldsymbol{\lambda}^{(2, \alpha)}$
are sufficient for decomposition of the two-electron part of the electronic Hamiltonian. 
Additional simplification came from the electron-spin 
symmetry in $g_{pq,rs}$ that allowed us to manipulate with orbitals ($N/2$) rather than spin-orbitals ($N$). Therefore, 
we used the unitary operators $\hat U_\alpha$ generated by exponentiation of $\LA{so}(N/2)$ instead of 
$\LA{u}(N)$, and our $\hat U_\alpha$'s were in the $Spin(N/2)$ subgroup of the original $U(N)$. 
As discussed in Ref.~\citenum{huggins2019efficient}, $\hat U_\alpha$ can be efficiently implemented 
($N^2/4-N/2$ two qubit gates and gate depth of exactly $N$) on a quantum computer with a limited connectivity. 
Limiting $\boldsymbol{\lambda}^{(2,\alpha)}$ to real entries and accounting for its symmetric property 
($\lambda^{(2,\alpha)}_{tu} = \lambda^{(2,\alpha)}_{ut}$), the FRO procedures had $N^2 / 4$ parameters in total for each fragment. For all molecules but H$_2$ parameters $w$ (see \eq{eq:IVFROcost}) and $\mu$ are set to $0.5$ and $30$, for $\rm H_2$, $w = 0.5$ and $\mu = 1$. To obtain fragments variance 
estimates for the VGFRO optimization we used Hartree-Fock wavefunctions.

As expected, the QWC method results in one of the highest numbers of measurable groups and one of the highest numbers of needed measurements. 
The FRO method without any concern about fragment 
variances provides the minimum number of fragments (Table~\ref{tab:ngroup_result}).  
However, more measurable fragments does not necessarily mean more measurements. 
As shown in Tables~\ref{tab:ngroup_result} and \ref{tab:variance_result}, the GFRO and VGFRO methods partition the Hamiltonians into considerably more measurable parts than SVD and FRO, 
yet they improve on SVD consistently by 10-30 percent in the number of measurements. 
%and all fermionic-algebra methods are up to ten times more efficient than the QWC method. 
This confirms the efficacy of the heuristic that groups operators with large coefficients in the same measurable fragment. 

Interestingly, for the smaller molecules, FC-SI outperforms the fermionic-algebra methods.
This suggests that the CSA decomposition based on commutativity between Pauli products in qubit-algebra is better at collecting large coefficients in few groups.
However, since the number of measurable fragments in FC-SI and SVD correspondingly scale with $O(N^3)$ and $O(N^2)$\cite{ChicagoA, huggins2019efficient}, we expect that as the size of the systems grows, the fermionic-algebra methods that have more flexible measurable parts will better capture most coefficients in the Hamiltonians with fewer fragments, 
and thus result in lower number of measurements due to the nature of \eq{eq:sum_sqrt}. 
The results for $N_2$ confirm that the advantage of FC-SI over SVD is shrinking, and the GFRO and 
VGFRO outperform FC-SI.

\section{Conclusions} \label{conclusion}

In this work we provided a unifying framework for many recently suggested approaches of efficient
partitioning of quantum operators into measurable fragments. The framework is based on identifying 
a Lie algebra that is used for the operator expression (embedding), analysis of the 
corresponding Lie group action, and the Cartan sub-algebra of the Lie algebra. 
The latter encodes the measurable fragments since it involves mutually commuting terms. 

To obtain measurable fragments we suggested two types of unitary transformation. 
First, the unitaries that are elements of the Lie group corresponding to the Lie algebra, \eq{eq:Ua}. 
These unitaries have the advantage of conserving the degree of the Lie algebra 
polynomials in the operator expression. Second, the unitaries that preserve the number of terms, 
\eq{eq:U2}. These unitaries use associative multiplication properties of algebraic operators 
and for qubit algebras correspond to the Clifford group. An intuitive rule to select between 
these choices of unitaries is the degree of the Lie algebraic operator polynomial expression: 
the Lie group unitaries are more efficient for lower degrees, and the number of term conserving 
unitaries are more useful for higher degrees.   

Being able to embed a single operator in multiple Lie algebras opens directions for further 
search for efficient partitioning schemes. Here, 
we mainly focused on embedding of the electronic Hamiltonian in fermionic ($\LA{u}(N)$) and qubit 
($\LA{su}(2^N)$) Lie algebras. 
%As was seen from results of previous works on partitioning of the electronic Hamiltonian, the fermionic embedding provides more efficient schemes than those of the qubit embedding. 
%Yet, this trend can change if the operator of interest would come from spin-related properties.
While the qubit embedding provides a more efficient scheme for electronic Hamiltonians of fewer orbitals 
than those of fermionic embedding, this trend can change as the size of the system grows. 

To minimize the total number of measurements, it is not enough to reduce the number of the measurable fragments
because of possible increase of fragments variances. It was found that grouping algorithms that employ 
greedy techniques were advantageous for lowering of the overall variance of the expectation value estimator. This 
advantage can be attributed to lowering the total variance in cases when fragment variances are distributed 
non-uniformly. In addition, the fermionic CSA decomposition allows one to optimize partitioning by 
using fragment variance estimates and to lower the overall variance in VGFRO further than 
in the greedy approach (GFRO).     

%The new framework allowed us to find a more efficient partitioning scheme based on fermionic embedding. 
%For several test systems, the new scheme consistently outperforms the previously most efficient scheme reported in Ref.~\citenum{huggins2019efficient}
%using the same type of unitary transformations. 
%\\ 

\section*{Acknowledgements}
 A.F.I. is grateful to Nicholas Rubin and William J. Huggins for providing details on performance of the algorithm from 
 Ref.~\citenum{huggins2019efficient} and acknowledges financial support from the Google Quantum Research Program, 
 Early Researcher Award and the Natural Sciences and Engineering Research Council of Canada.

\section*{Appendix A: Various embeddings of the electronic Hamiltonian}

The electronic Hamiltonian can be written as 
\bea\label{eq:H2} 
\hat H_e = \sum_{pq} h_{pq} \EE{p}{q} + \sum_{pqrs} g_{pq,rs} \EE{p}{q} \EE{r}{s}
\eea
where $h_{pq}$ and $g_{pq,rs}$ are one- and two-electronic integrals that are real constants,\cite{Helgaker} 
and $\EE{p}{q} = \CR{p}\AN{q}$ 
are elements of the $\LA{gl}(N)$ Lie algebra\cite{Gilmore:2008}
\bea
[\EE{p}{q},\EE{r}{s}] = \EE{p}{s}\delta_{qr} - \EE{r}{q}\delta_{ps}.
\eea 
$\CR{p},\AN{q}$ operators are regular fermionic creation and annihilation operators where $p,q$ run over $N$ 
spin-orbitals. The CSA for $\LA{gl}(N)$ consists of $N$ operators $\EE{p}{p}$.  
A more common second quantized form of $\hat H_e$  
\bea \label{eq:H1}
\hat H_e = \sum_{pq} \tilde{h}_{pq} \CR{p}\AN{q} + \sum_{pqrs} \tilde{g}_{pq,rs} \CR{p}\CR{q}\AN{r}\AN{s}
\eea
can be considered as a different embedding (here $\tilde{h}_{pq}$ and $\tilde{g}_{pq,rs}$ are real constants). 
To close the Lie algebra containing $\CR{p}$ and $\AN{q}$ operators, one also needs to add products
$\{\CR{p}\AN{q}, \CR{p}\CR{q}, \AN{p}\AN{q}\}$, which makes the $\LA{so}(2N+1)$ Lie algebra.\cite{Fukutome:1981/65} 

%%%%%%%%%%%%%%%%%%%%

Another class of embeddings can be obtained by mapping fermionic operators to qubits using 
the Jordan-Wigner (JW), Bravyi-Kitaev (BK), or similar fermionic-qubit mappings.
\cite{Bravyi:2002/aph/210, Seeley:2012/jcp/224109,Tranter:2015/ijqc/1431,Setia:2017/ArXiv/1712.00446,Havlicek:2017/pra/032332} Here, we will use the JW mapping as the simplest for illustrative purpose
\bea
\AN{p} &=& (\hat x_p-i\hat y_p)\otimes \hat z_{p-1} \otimes \hat z_{p-2} ...  \otimes \hat z_{1} \\
\CR{p} &=& (\hat x_p+i\hat y_p)\otimes \hat z_{p-1} \otimes \hat z_{p-2} ...  \otimes \hat z_{1}.
\eea
This mapping produces 
\bea
\hat H_q = \sum_{k} c_k \hat P_k,
\eea
where $c_k$ are numerical constants, and $\hat P_k$'s are defined in \eq{eq:Oq}. 
We can consider $\hat P_k$'s as elements of the UEA where the Lie algebra is a direct sum 
of $N$ $\LA{su}(2)$'s: $\mathcal{S} = \LA{su}(2)\oplus ... \oplus \LA{su}(2)$ and $\mathbb{K} = \mathbb{R}$. 
$\mathcal{S}$ is a semisimple Lie algebra with $3N$ generators ($\hat x_k, \hat y_k, \hat z_k$). 

Alternatively, instead of the UEA one can consider its isomorphic algebra, $\LA{su}(2^N)$ with $4^N-1$ generators 
given by the $\hat P_k$ operators.  
Accounting time-reversal symmetry of the electronic Hamiltonian makes all $c_k$'s real and all $\hat P_k$'s to 
contain only products of even number of $\hat y_k$ operators, hence, $\hat H_q$ is an element of the 
$\LA{so}(2^N)$ sub-algebra of $\LA{su}(2^N)$. 
Thus, the qubit embedding $\hat H_q$ can be either seen as an element of the UEA of $\mathcal{S}$ or 
an element of the $\LA{so}(2^N)$ Lie algebra. 

There are $3^N$ CSAs for $\mathcal{S}$, which are based on selecting a particular 
Pauli operator ($\hat x, \hat y$ or $\hat z$) for each qubit and thus containing $N$ elements each, 
for example, a directly measurable CSA is $\{\hat z_k\}_{k=1}^N$. \\

Out of the total $2^N(2^N-1)/2$ generators of $\LA{so}(2^N)$, one can select exponentially many 
CSAs. They can be obtained by constructing UEAs from CSAs of $\mathcal{S}$ and removing those 
that contain an odd number of $\hat y$ operators 
to maintain a real character. This generates exponentially large CSAs with $2^N-1$ commuting elements,
for example, for two qubits a directly measurable CSA has basis elements $\{\hat z_1,\hat z_2,\hat z_1\hat z_2\}$. 

\section*{Appendix B: Singular Value Decomposition for the two-electron Hamiltonian}

Here we show how SVD can provide a heuristic factorization of the two-electron Hamiltonian part
\bea
\hat H_e^{(2)} = \sum_{pqrs}^N g_{pq,rs} \EE{p}{q} \EE{r}{s}.
\eea
SVD for $g_{pq,rs}$ gives
\bea
g_{pq,rs} &=& \sum_k^{N^2} U_{pq,k} \Lambda_k U^\dagger_{k,rs}  \\
&=& \sum_k^{N^2} U_{pq,k} \Lambda_k^{1/2} [U_{rs,k} \Lambda_k^{1/2}]^\dagger \\
&=& \sum_k^{N^2} L^{(k)}_{pq} (L^{(k)}_{rs})^\dagger,
\eea
where $\Lambda_k$ are singular values. 
Each matrix $L^{(k)}_{pq}$ can be diagonalized as  $\sum_r(U^{(k)})^\dagger_{pr} \omega_r^{(k)} U^{(k)}_{rq} $ for fixed $k$, 
this allows us to write the two-electron tensor as   
\bea\notag
g_{pq,rs} &=& \sum_{k}^{N^2}\sum_{tu}^N (U^{(k)})^\dagger_{pt} \omega_t^{(k)} U^{(k)}_{tq}  (U^{(k)})^\dagger_{ru} \omega_{u}^{(k)} U^{(k)}_{us}. 
\eea\\
The entire $\hat H_e^{(2)}$ then can be written as
\bea \notag%\label{eq:kpart}
\hat H_e^{(2)} &=& \sum_{pqrstu, k}  [(U^{(k)})^\dagger_{pt} \EE{p}{q}  U^{(k)}_{tq}]  \omega_t^{(k)}  \omega_u^{(k)} [(U^{(k)})^\dagger_{ru} \EE{r}{s} U^{(k)}_{us}].
\eea
It is easy to show that the operators in the square brackets
\bea
\hat B_t^{(k)} = \sum_{pq} (U^{(k)})^\dagger_{pt} \EE{p}{q}  U^{(k)}_{tq}
\eea
are from the CSA and are equivalent to $\EE{t}{t}$.

For showing commutativity of $\hat B_t^{(k)}$ let us consider their commutators for a fixed $k$ 
\bea\notag
~[\hat B_t^{(k)},\hat B_u^{(k)}] &=&  \sum_{pq,rs} (U^{(k)})^\dagger_{pt} U^{(k)}_{tq} (U^{(k)})^\dagger_{ru}U^{(k)}_{us} 
[\EE{p}{q},\EE{r}{s}] 
\eea
$[\EE{p}{q},\EE{r}{s}] = \EE{p}{s}\delta_{qr} - \EE{r}{q}\delta_{ps}$, therefore we can consider two parts
\bea
\EE{p}{s}&:& ~\sum_{pq,rs} (U^{(k)})^\dagger_{pt} U^{(k)}_{tq} (U^{(k)})^\dagger_{ru}U^{(k)}_{us} \EE{p}{s} \delta_{qr}\\
&=&  \sum_{pqs} (U^{(k)})^\dagger_{pt} U^{(k)}_{tq} (U^{(k)})^\dagger_{qu}U^{(k)}_{us} \EE{p}{s}\\
&=&  \sum_{ps} (U^{(k)})^\dagger_{pt} \delta_{tu} U^{(k)}_{us} \EE{p}{s} \\
&=&  \sum_{ps} (U^{(k)})^\dagger_{pt} U^{(k)}_{ts} \EE{p}{s} = \hat B_t^{(k)}. 
\eea
Analogously, for the $\EE{r}{s}$ part we will obtain the same result but with the minus sign, thus the commutator is zero.

Therefore, one can measure each $k$ part of the two-electronic Hamiltonian 
\bea 
\hat H_e^{(2)} &=& \sum_k^{N^2}\sum_{tu}^N \omega_t^{(k)} \omega_u^{(k)} \hat B_t^{(k)} \hat B_u^{(k)}.
\eea 
Considering the norm of $\Omega_{tu}^{(k)}=\omega_t^{(k)} \omega_u^{(k)}$ for each $k$, 
one can reduce the number of measurable
sets if this norm is below a certain threshold.

\section*{Appendix C: Details of Hamiltonians}

The Hamiltonians were generated using the STO-3G basis and the BK transformation. The nuclear geometries for the Hamiltonians are 
R$(\rm H-H) = 1 \AA$  ($\rm H_2$), R$(\rm Li-H) = 1 \AA$ ($\rm LiH$), R$(\rm Be-H) = 1 \AA$ with collinear atomic arrangement ($\rm BeH_2$), 
R$(\rm O-H) = 1 \AA$ with $\angle HOH = 107.6^\circ$ ($\rm H_2O$), 
R$(\rm N-H) = 1 \AA$ with $\angle HNH = 107^\circ$ ($\rm NH_3$), 
and R$(\rm N-N) = 1 \AA$ ($\rm N_2$).

%\bibliographystyle{apsrev4-1}
%\bibliography{library}

%merlin.mbs apsrev4-1.bst 2010-07-25 4.21a (PWD, AO, DPC) hacked
%Control: key (0)
%Control: author (72) initials jnrlst
%Control: editor formatted (1) identically to author
%Control: production of article title (-1) disabled
%Control: page (0) single
%Control: year (1) truncated
%Control: production of eprint (0) enabled
%

\end{document}